# On X-ray scattering model for single particles Part I: The legacy of protein crystallography

Aliakbar Jafarpour

*Dept. of Biomolecular Mechanisms, Max-Planck Inst. for Medical Research, Jahnstr 29, 69120 Hiedelberg, Germany*
*jafarpour.a.j@ieee.org*

**Abstract:** Emerging coherent X-ray scattering patterns of single particles have shown *dominant morphological* signatures in agreement with predictions of the scattering model used for conventional protein crystallography. The key question is if and to what extent these scattering patterns contain *volumetric* information, and what model can retrieve it. The scattering model of protein crystallography is valid for very small crystals or those like crystalized biomolecules with small *coherent subunits*. But in the general case, it fails to model the integrated intensities of diffraction spots, and cannot even find the size of the crystal. The more rigorous and less employed alternative is a purely-classical crystal-specific model, which bypasses the fundamental notion of *bulk* and hence the non-classical X-ray scattering from bulk. This contribution is *Part 1* out of two reports, in which we seek to clarify the assumptions of some different regimes and models of X-ray scattering and their implications for single particle imaging. In this part, the predictions of the conventional and the rigorous models for emerging scattering patterns of protein nanocrystals (intermediate case between conventional crystals and single particles) are contrasted, and the terminology conflict regarding "Diffraction Theory" is addressed. With a clearer picture of crystal scattering, *Part 2* will focus on additional concepts, limitations, correction schemes, and alternative models relevant to single particles. Aside from such optical details, protein crystallography is an advanced tool of analytical chemistry and not a self-contained optical imaging technique (despite significant instrumental role of optical data). As such, its final results can be neither confirmed nor rejected on mere optical grounds; i.e., no *jurisdiction* for optics.

## 1. Introduction

### 1.1. Context of the problem

Conventional X-ray crystallography has an unparalleled capability and record in atomic-resolution 3D imaging of crystalized biomolecules [1]. The recent efforts seeking to extend this technology to nanocrystals and eventually single particles [2] have been listed as one of the 10 breakthroughs of the year by the Science Magazine [3]. These efforts have resulted in appropriate alternatives for the technology behind crystallography in most aspects: sample preparation, sample delivery, light source, optoelectronic detection, and statistical analysis. It seems however, that the (over-) simplified *model of light propagation* and its crucial implicit assumptions have been taken in many (if not most) cases for granted. This model is not directly applicable to arbitrary illumination schemes or arbitrary heterogeneous objects (even arbitrary crystals).

The validity of the scattering model of conventional protein crystallography (plane wave propagation inside a crystal) is based on partial coherence of light scattering. According to an intuitive model depicted in Fig. 1, coherent scattering from small perfect subunits (*mosaic blocks*) are added incoherently [4]. The total scattered intensity pattern resembles that of individual subunits despite variations of the illumination that each block receives.

Most protein crystallography studies are only concerned with imaging an asymmetric subunit of the unit cell and not the entire object. The incident light is confined to the interior of the entrance facet of the crystal, and the sizes and relative orientations of different subunits, with some of them not even being illuminated, are not relevant. However, a simplified version and a common perception of this story is "*uniform coherent* illumination of the *entire object*", attributed to the (important, yet exaggerated) role of weak light-matter interaction. This simplified picture of light propagation for the end-users and even software-developers is the outlook of a well-engineered technology (with special considerations in illumination, imposing specific symmetry constraints, model-based fit and refinement of the retrieved 3D profile, having meaningful figures of merit …), which has revolutionized structural biology by identification of several structure-function biological phenomena [1].

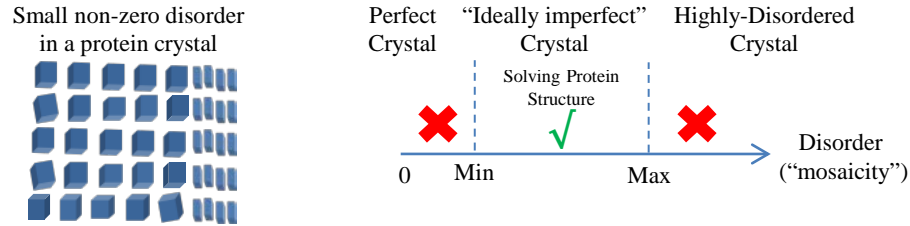

Fig. 1. (Left) A protein crystal has a small yet nonzero level of disorder; intuitively modeled by slightly-disoriented mosaic blocks, which are represented here as small cubes. (Right) Optically, the simple scattering model employed in protein crystallography makes sense when a crystal is neither perfect, nor highly-disordered, as it seems to be the case in protein crystals. With this ideally imperfect structure, a simple model of light scattering seems to be efficient enough to do volumetric 3D imaging of crystallized proteins at record (atomic) resolution. Ironically, much more tedious and rigorous scattering models, applicable to perfect crystals, "fail" (Sec. 2.3). Emerging scattering patterns of single particles and even nanocrystals seem to change this picture and the relevance of alternative scattering models.

As shown in Section 3, basic properties of nanocrystals (size of the crystal, route to ordinary crystals, and the dependence of boundary fringes on material properties) cannot be found correctly with the conventional model of protein crystallography. However, an acceptable 3D reconstruction (by elegant use of extensive level of meaningful *a-priori* information, especially with the Molecular Replacement technique) is possible. While insightful for developing correction schemes in nanocrystallography, such counter-intuitive observations have more serious messages for *coherent* single-particle reconstruction and induce second thoughts about the perceived legacy inherited from protein crystallography.

*1.2. Different meanings of "3D imaging": volumetry vs. morphology*

3D imaging in protein crystallography or its macroscopic counterpart, computed tomography (CT-scan), refers to the retrieval of a full 3D *volume density* function in a given region as $D = f(r, \theta, \phi)$, where the arguments of the function are spherical coordinate variables. Contrary to such volumetric studies, a *morphological* study finds a *surface* function (of a shell or a homogeneous object).

The information content of volumetric imaging is modeled with a 3D function, whereas that of a morphological study is a pair of 2D functions (images) modeling the densities and radial distances of the surface points $D = d_{surf}(\theta, \phi)\delta[r - r_{surf}(\theta, \phi)]$ for a shell. Morphological information of a homogeneous (shell or solid) object is just a 2D image $r_{surf}(\theta, \phi)$, even though it pertains to an object embedded in 3D space. Many "3D imaging" projects in computer graphics, for instance, need to retrieve morphological information about opaque object(s) embedded in an otherwise-empty scene (not a full non-sparse map of volume density).

Aside from this mathematical distinction, the physical interaction between an object and a light beam may happen just at the boundaries or in the entire volume of the object down the propagation path. In volumetry, the light beam should maintain its key properties (coherence, directionality …) while passing through the entire volume, or its variations should be accounted for in image reconstruction. These requirements can be significantly relaxed in the case of surface morphology.

Morphological studies with a given technology can be important steps towards, but not proof of the concept for volumetry with that technology.

*1.3. Definition of the problem*

The block diagram of the inverse problem of X-ray scattering from single particles is shown in Fig. 2. The Figure shows the significance of the scattering model in estimating or validating a candidate solution and its impact on the reconstruction process.

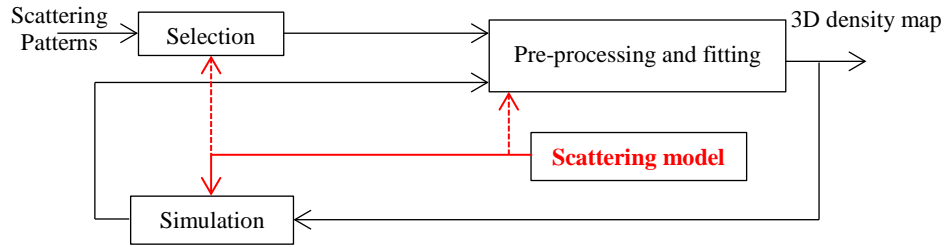

Fig. 2. Simplified block diagram of an optimization-based 3D reconstruction of a density map of a *single particle* from measured scattering patterns. Simulation of scattering patterns and also the selection of "good" scattering patterns and filtering "distortions" out are affected by the choice of the scattering model. Part of the ambiguity in assigning a 3D structure to a series of scattering patterns originates from inaccuracies of the scattering model.

According to the scattering model of protein crystallography, at large (classical) scales, the pattern of scattered light from a homogeneous object is material-independent. The pronounced signatures of morphology may confirm this prediction in many cases in a qualitative or semi-quantitative way. However, it is clear that light undergoes more distortions in one material compared to another, and hence the constraint on the validity of the model itself is material-dependent.

Experimental scattering patterns from symmetric nanoparticles [5] or viruses [6] have shown such major signatures of morphology. The key question is if the weak residual patterns have also some volumetric information; and if yes, what constraints exist for the retrieval.

Given the common use of the scattering model of protein crystallography for single particles, clarification of crystal scattering models (and their scopes of validity) covers part of the puzzle of scattering models of single particles, as depicted in Fig. 3.

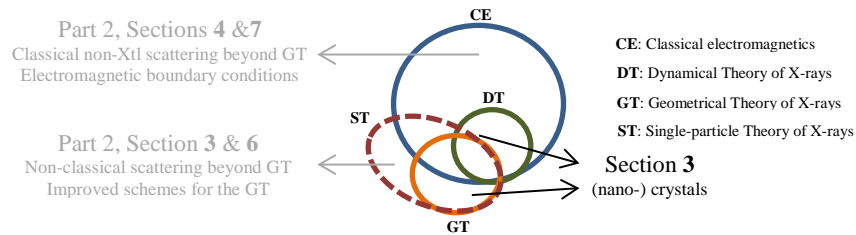

Fig. 3. Venn diagram showing different regimes of X-ray scattering and the interrelation between them. It is the big picture, within which the interrelation between the materials in subsequent Sections (and also those in *Part 2* of this report, which have been grayed out here) can be seen.

*1.4. Outline*

A concise and simplified overview of existing scattering models is presented in Section 2. Predictions of the conventional and the alternative scattering models for the emerging case of protein nanocrystals are discussed and contrasted in Section 3. Terminology conflicts and correspondingly potential misconceptions are addressed in Section 4, with focus on common out-of-context use of Diffraction Theory to model volumetric scattering. More aspects of scattering in crystals and the differences between protein crystallography and an optical imaging technique are further discussed in Section 5, and conclusions are made in Section 6. Finally, Appendices A-D present more details of the materials presented in earlier Sections.

**2. Basic concepts in X-ray scattering**

*2.1. Different regimes of light-matter interaction*

The term "bulk" in X-ray literature is a *macroscopic* concept [7]. At non-classical sizes (below $\sim 10nm$) or resolutions (above $\sim 1/(40nm)$), *bulk* may lose its key electromagnetic properties in 1) trivial association with spatial map of different materials, or 2) supporting a sustained plane wave as an electromagnetic mode. The distinction between classical and non-classical scattering is crucial for single particles and is detailed in *Part 2*.

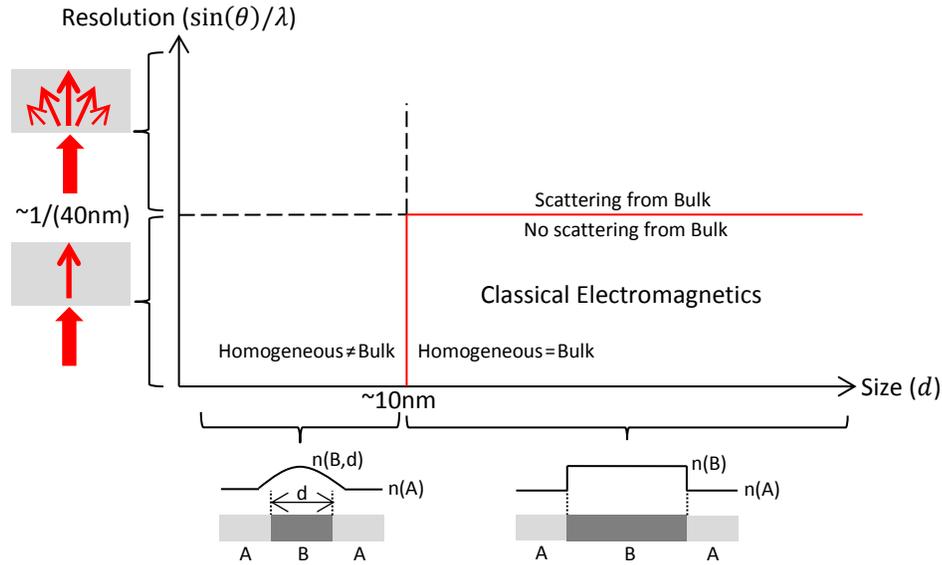

Fig. 4. The boundaries of the notion of *bulk* and *direct* applications of classical electromagnetics. The cartoons underneath the horizontal axis show that at small scales, a chemically-homogeneous material does not necessarily have a uniform or even confined electron cloud. The cartoons along the vertical axis show that at high resolutions, a plane wave is not a sustained electromagnetic mode of a homogeneous medium and causes scattering.

*2.2. Uniformity of internal illumination and three models*

Far-field scattered light can be traced back to the spatial profile of an *illuminated density*, as shown in Fig. 5. In a (semi-) classical picture, the weak density at X-ray wavelengths is proportional to the contrast in dielectric constant, which is also the refractive index contrast within a factor of two: $\rho = 1 - \epsilon_r = 1 - n^2 = 1 - (1 - \delta n)^2 \sim 2\delta n$. In addition to far-field scattered light, the density creates a second signature of itself in the *internal* illumination pattern; i.e., the local electric field inside the object in response to the external incident electric field. Non-uniform internal illumination in a homogeneous object can appear as a spurious gradient of density when using an approximate model of far-field scattering (with assumption of uniform internal illumination).

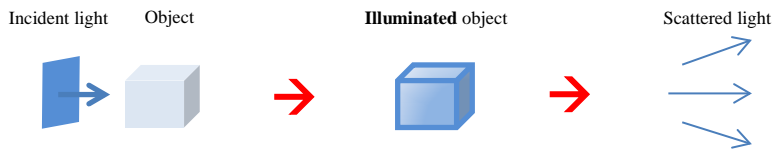

Fig. 5. The light incident on an object creates not only a scattering pattern, but also an illumination pattern inside the object. This *internal* illumination pattern and the dynamics of its formation have an implicit signature in the scattering pattern. A scattering model may simply assume that the incident light sweeps the object without being distorted (*Geometrical* model), or rigorously calculate the internal illumination and the scattered light as *coupled* unknowns (*Electromagnetic* model), or model the nontrivial internal illumination, but with the simplifying assumption that it is disentangled (and can be estimated independently) from the scattered light (*Hybrid* models), as implied by this Figure.

The internal illumination (local field) in a scattering object is addressed in three models, as follows:

*1. Geometrical model*: The local field is assumed to be the same as the incident field ("weak light-matter interaction"). Assuming an incident plane wave, the local field will be uniform in amplitude across the scattering volume, and its propagation phase is absorbed in the geometrical distribution of the density. This model, also referred to as the Kinematic model, is used in the formulation of X-ray scattering in protein crystallography.

*2. Classical Electromagnetic model*: The coupled scattered and local fields are both found simultaneously by satisfying classical Maxwell's equations (including boundary conditions). With this model, non-classical effects of X-ray scattering from bulk or coupled evolution of optical field and electronic wavefunction, if any and if not accounted for otherwise, are discarded. In the case of small low-contrast scattering objects, the Electromagnetic model is reduced to a form similar to the Geometrical model (Rayleigh-Gans or the first-order Born's approximation).

*3. Hybrid models*: A Hybrid model goes beyond the Geometrical model and accounts for non-uniform illumination, yet with the simplifying assumption of disentangled illumination and scattering. In a simple form, a smoothly-depleted plane wave may be assumed. In a more rigorous approach for an object with known boundaries and macroscopic bulk index, first the local field is estimated *classically* (Fresnel-like reflections off the boundaries) by explicit consideration of (planar) boundary conditions. The unknown embedded density is then considered as a perturbation coupling two *eigenstates* in a $1^{st}$-order perturbation scheme [8].

Some different contexts of applications of these three models of X-ray scattering have been briefly listed in Appendix A. It is noted specifically that non-uniform internal illumination can have counter-intuitive implications (Appendix B) for the damage process, which is an important concern in X-ray studies.

*2.3. Mosaicity model: Transition from the Electromagnetic to the Geometrical model*

In the early years of X-ray crystallography, both the Geometrical and (a simpler version of) the Electromagnetic model were tested. Both models were similar in the sense of predicting the existence and the positions of diffraction spots; namely the discrete 3D Fourier harmonics of the density. The main question was the (integrated) *intensities* of these spots, which is also the main question for recent nanocrystallography experiments [9]. Experimental data agreed with the prediction of the simple Geometrical model ($I \propto |SF|^2$) and not the prediction of the more rigorous Electromagnetic model ($I \propto |SF|^1$; for a *perfect* crystal) [10]. Here $I$ is the integrated intensity of a diffraction spot, and $SF$ (structure factor) is proportional to the 3D Fourier transform of the (uniformly illuminated) density, calculated at the reciprocal space point corresponding to the diffraction spot.

These observations were explained by the existence of incoherence in crystals and the notion of *mosaicity* [10], as shown in Fig. 1. Many inorganic and most protein crystals are *not* perfect. A hypothetical cross section of the object can be thought of as showing a wall formed with many small mosaic blocks; i.e., perfect crystals with small relative disorientation. Said differently, with the electromagnetic model, an approximate description of the object and a rigorous description of light propagation (in the wrong object) give an incorrect answer. With the Geometrical model, an approximate description of the object and an approximate estimation of illumination give the correct answer ("two wrongs make a right"). Fig. 6 shows the reduction of the 2-beam model to the Geometrical model in a small crystal or in a protein crystal with small coherent subunits (*mosaic blocks*).

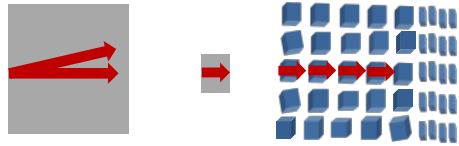

Fig. 6. (Left) Generation of two intense beams (not weak scattered rays) inside a perfect crystal. (Middle) In the thin crystal regime, there is nearly no distinction between the two beams. (Right) Propagation of a single beam inside coherent sub-units of a protein (ideally imperfect) crystal, as opposed to two beams in the case of a perfect crystal. The single beams inside individual sub-units do not have to be the same, as shown in Fig. 13 (Appendix B).

### 3. Nanocrystals

Contrary to conventional protein crystallography, recent experiments [11] are performed 1) on multiple crystals with random sizes and orientations delivered via a liquid jet, rather than on a single mounted rotating crystal, and 2) with the crystals *immersed in a large-diameter* beam, rather than the beam being confined to the interior of the input facet.

Here we review the implications of the Geometrical and the Electromagnetic models for perfect crystals in the two cases of infinite and finite transverse dimension. We then address experimental data briefly, and draw some conclusions for improving the common (Geometrical) model.

*3.1. Geometrical model*

According to the Geometrical model, the effect of the finite size of an $N$-th paralellpiped crystallite is simply the introduction of a modulation factor, referred to as shape transform: $I(q_x, q_y, q_z) = I_{uc}(q_x, q_y, q_z) Sh(\tilde{q}_x, \tilde{q}_y, \tilde{q}_z)$, where $I_{uc}$, $I$, and $\boldsymbol{q} = (q_x, q_y, q_z)$ are the scattered intensity of a unit-cell, the scattered intensity of the crystal with $N_1 \times N_2 \times N_3$ unit-cells, and the scattering vector, respectively. The vector $\tilde{\boldsymbol{q}} = 2\pi[\boldsymbol{a}|\boldsymbol{b}|\boldsymbol{c}]^T \boldsymbol{q}$ depends on the projections of the scattering vector along the crystal axes $\boldsymbol{a}$, $\boldsymbol{b}$, and $\boldsymbol{c}$. The shape transform, sketched in Fig. 7, originates from shift-related complex exponentials and equals $Sh(q_x, q_y, q_z) = \left( \frac{\sin(N_1 \tilde{q}_x)}{\sin(\tilde{q}_x)} \frac{\sin(N_2 \tilde{q}_y)}{\sin(\tilde{q}_y)} \frac{\sin(N_3 \tilde{q}_z)}{\sin(\tilde{q}_z)} \right)^2$ [9]. Modulations, as predicted by the shape transform of the Geometrical model have been observed in nanocrystals; see Figure 2 in [11].

These implications of the shape transform are noteworthy:
- Plane-wave approximation; i.e., the crystal being immersed in a beam with a diameter larger than the size of the input facet of the crystal (inconsistent with the illumination scheme in the established case of protein crystallography if $N_{1,2,3}$ represent the size of the *crystal*; consistent if $N_{1,2,3}$ represent the size of a *mosaic block*)
- No difference if the beam enters the crystal from one or multiple facets (aside from a rotation of coordinate)
- Indifference with respect to the direction of light propagation or normal to it

- Fringe spacing merely depending on geometry and not material property: $N_1 \tilde{q}_{x,y,z} = n\pi$
- Fringes oriented along the crystallographic axes $\boldsymbol{a}, \boldsymbol{b}, \boldsymbol{c}$ as a separable function of $(\tilde{q}_x, \tilde{q}_y, \tilde{q}_z)$
- Approaching Dirac's delta function at the position of diffraction spots and disappearance of fringes, as $N_1, N_2, N_3 \to \infty$ and a gain factor proportional to $(N_1 N_2 N_3)^2$

In conventional protein crystallography, modulations as predicted by the shape transform are not common. However, as a crystal is rotated, it first approaches and then deviates from the condition of Bragg resonance (for a given diffraction spot). The finite spread of a diffraction spot in the scattering coordinate as a function of this detuning angle is referred to as the *Rocking curve* [12], as shown in Fig. 7. The inverse of its width is a *direct* measure of incoherence ("mosaicity").

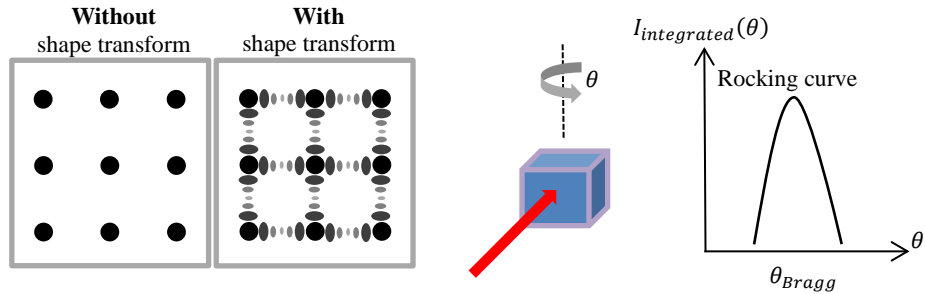

Fig. 7. (Left) When the incident light sees the finite transverse size of a crystal, modulations around the diffraction spots, referred to as the Shape Transform, may be observed. (Middle) As the crystal is rotated around an axis perpendicular to the direction of beam propagation, (Right) one obtains the Rocking curve, which shows the integrated intensity of a specific diffraction spot as a function of the detuning angle. Both shape transform and rocking curve show the spread of a diffraction spot in the scattering coordinate.

A bell-like (Gaussian, cosine ...) model of the rocking curve is also a crucial part of conventional data analysis of protein crystals [13-15]. The best fit of all diffraction spots to a 3D density map also determines the best *indirectly* measured mosaicity. In the limit of zero mosaicity, the *model* rocking curve of data processing approaches a Delta function. It is sometimes referred to as the "ideal" case, from which point one may start to study the effects of crystal imperfections. However, the entire theoretical foundation of all these formulations is based on a finite nonzero mosaicity, as shown in Fig. 1. Irrespective of numerical patterns and the possibility of good fits, *the model rocking curve is in principle invalid, when extrapolated to zero or too small values of mosaicity*. Finding patterns in this regime [16] can be insightful.

Despite differences, both the shape transform and the rocking curve are important coherent signatures of a crystal. In practice, their estimated models determine how to interpolate/convert a measured distribution of light intensity around a diffraction spot into a single value (which would correspond to an "ideal" crystal). This crucial integration will be further discussed in Sec. 3.3.3 and 5.3.

### 3.2. Electromagnetic model ("Dynamical Theory")

At X-ray wavelengths and in a *perfect* crystal with a cross section larger than the incident light, the entrance facet generates two (intense) refracted *beams*, which generate many (additional low-intensity) *k vectors*. Fresnel-like reflections off the exit facet further change the local field by forming a standing wave pattern with two pairs of counter-propagating *beams* (for a given polarization). These experiments may be done in the *Laue regime* (transmission geometry) or *Bragg regime* (reflection geometry). However, when the illumination sees the transverse edges of the object, even conventional forms of Electromagnetic theory cannot be used. Multiple reflections of *beams* (not scattered photons) give rise to mixed regimes such as Laue-Bragg [17] and Laue-(Bragg)$^M$ [18]. Reflection of polyhedral crystals [19], the intensity patterns [20] and rocking curves [21] of a crystal with finite cross section, and the effect of partial illumination of a facet [22] have also been formulated. More details about Electromagnetic description of scattering in crystals can be found in Appendix C.

The key differences between Electromagnetic and Geometrical models are related to the total intensity and the spread of a diffraction spot (shape transform or rocking curve), tabulated in Table 1. We note specifically that in the Electromagnetic picture, the fringe spacing around a diffraction spot depends not only on the size, but also on the value of the structure factor [10,23]. As such, the size of the "coherent subunit" of a crystal (not to mention the size of the entire crystal) is not related to fringe separations in a trivial way.

Table 1: Contrasting the predictions of the Geometrical and the Electromagnetic model for nanocrystals

| Model | Material-dependent fringes | Number of entrance facets | Route to ordinary crystals | Trivial size estimation | $\int I dq$ vs. $N$ |
|---|---|---|---|---|---|
| Geometrical | **No** | **Unimportant** | $N \to \infty$ | **Yes** | $\to \infty$ |
| Electromagnetic | **Yes** | **Important** | $1 \leftarrow N$ | **No** | **Oscillatory** |

*3.3. Experimental results and discussions*

*3.3.1. Nanocrystals seem to have larger "mosaic blocks"*

A key point in understanding nanocrystals is the route from their *coherent* signatures to those of conventional crystals. According to the Geometrical model, even with very few unit cells, one sees the oscillations of the sinc-like shape transform. The oscillations disappear, as the number of unit cells [*per mosaic block*] approaches infinity. In such a regime, however, the Geometrical model fails, as it is valid for small mosaic blocks. This is not the route to conventional crystals.

According to the Electromagnetic model, for perfect crystals with infinite cross section, passing the critical length in the transmission geometry has a clear signature of developing oscillations (with nearly equal amplitudes and possibly with a minimum at Bragg condition) in the rocking curve [10]. In other words, a necessary condition for the validity of the Geometrical model in this case is the lack of oscillations in the rocking curve. Interestingly, once the oscillations are observed (around a critical length), further increase of the thickness *increases* the fringe spacing in the rocking curves calculated for crystals with no absorption, as shown in Fig. 5.1.6.6 in [10]. This may also seem contradictory with the intuition developed with the Geometrical model.

The emergence of oscillations in rocking curve (with nearly equal amplitudes in the middle, as predicted by the Electromagnetic model) has also been reported for increasingly thicker *stable* protein crystals (= increasingly-thicker mosaic blocks) grown and characterized under minimal stress [12]. Under normal conditions with higher level of stress, however, increasingly thicker crystals will have increasingly more difficulty in accessing the surface and relieving their stress (switching from *soft* to *hard* mechanical boundary condition for inner unit-cells at the presence of stress and an increased chance of going from the elastic to plastic deformation regime). It is expected that under normal level of stress, increased size is associated with decrease in the size of the mosaic block. Equivalently, *smaller nanocrystals are likely to have larger coherent [mosaic] blocks, the coherent signatures of which are observed in scattering patterns*.

Even if the size of mosaic blocks remains the same in nanocrystals (and the scattering from individual blocks follows the Geometrical model), the reduced *number* of blocks can change the incoherence of superposition. A smaller ensemble of mosaic blocks implies less chance of uniform distribution of orientations and cancellation of cross terms $\langle I_{total} \rangle =$

$\langle |E_{total}|^2 \rangle = \langle \sum\sum E_m E_n^* \rangle = \sum I_n + \sum_{m \neq n} \langle E_m E_n^* \rangle \neq \sum I_n$, where $I$, $E$, and $\langle \rangle$ denote the intensity of scattered light, the electric field of scattered light, and the probabilistic expected value, respectively.

Increased size of coherent units and increased level of coherence in the superposition of scattered beams from such units introduce two challenges to the Geometrical model for the entire crystal. However, a more important question is whether the notion of "mosaic blocks" (used along with the Electromagnetic model in this discussion) is applicable to these cases, even for a semi-quantitative description of light propagation.

Shifted or modulated rocking curves also explain occasional lack of a specific diffraction spot $(h, k, l)$ in a well-indexed measured snapshot, where the simulated snapshot (according to the Geometrical model) predicts one.

### 3.3.2. Corrections to Geometrical model inspired by the Electromagnetic model

With the increased size of mosaic blocks, one ends up with a regime that neither the (conventional) Electromagnetic, nor the Geometrical model is directly applicable. More complicated versions of the Electromagnetic model may be interesting to develop. However, it is much more practical *and conclusive* for inverse problems to develop simple correction schemes for the Geometrical model inspired by the Electromagnetic model.

Despite the complex dependence of the rocking curve on longitudinal and transverse lengths for a given simple crystal and a given diffraction spot [21], the overall behavior may be simply represented as going from a Gaussian-like rocking curve towards one with asymmetry, modulations, and offset from the condition of Bragg resonance. All these behaviors may be simply modeled *phenomenologically* as in $I_{integrated}(\theta) = e^{-(\theta/\theta_1)^2[1+\theta/\theta_2]}[1 + m\cos(n\theta + \theta_3)]$. An insightful experiment will be one on a typical protein crystal with two illumination scenarios: 1) confined to the middle and 2) touching the edges of of the crystal. Direct measurements of rocking curves and also improving the global fits by optimizing the $\{\theta_1, \theta_2, m, n, \theta_3\}$ parameters provide insight into the impact of illumination on rocking curves, without adding additional degrees of uncertainty (size, crystal disorder …). A rocking curve may be split into 1) a common data-dependent part, and 2) a pre-determined data-independent $(h, k, l)$-dependent part.

For nanocrystallography, the fit parameters need to be found for individual snapshots, and alternative expressions for rocking curve (or shape transform) may be more helpful. The following observations may also be helpful in formulating an empirical rocking curve or shape transform:
- Possibility of reducing the 3D problem to a 1D one (based on the apparent separability of the shape transform observed in measured small-angle diffraction spots) as $Sh(\tilde{q}_x)Sh(\tilde{q}_y)Sh(\tilde{q}_z)$
- Simpler version of the Electromagnetic (Darwin's) theory for finite-size crystals [24]
- Selective computational suppression of coherence (in plain English: averaging a lot of data; explained in the next Sub-Section)

### 3.3.3. Averaging and unorthodox (modulated, asymmetric, shifted) rocking curves

For a given dataset, improving the accuracy of reconstruction may be only possible with a more accurate model for the rocking curve or shape transform. However, when possible, collecting as much data as one can may also give acceptable results with conventional (or simple modifications of the) models of shape transform or rocking curve. More data has two trivial benefits of 1) better sampling of the scattering space, and 2) better suppression of uncorrelated noise. The third nontrivial benefit seems to be creating a simple effective rocking curve for the intensity *averaged over snapshots* by enforcing a minor level of incoherence.

Averaging (*incoherent* intensity-wise superposition) of many properly-oriented (indexed) snapshots seems to wash out the *coherent* asymmetries, offsets, and modulations of the rocking curve [21], while preserving a reasonable (integrated) intensity to background ratio. Together with the *a-priori* information commonly superimposed in data processing and enforcing the Geometrical model, it may make the common symmetric bell-like rocking curve of the Geometrical model effective again. This third nontrivial effect of averaging may be a key contributor to the results shown in Fig. 4 in [25], which has reported the only nanocrystallography reconstruction with minimal use of *a-priori* information (with *de novo* phasing) to date.

Said differently, 3D merging of many 2D scattering patterns in conventional crystallography tries to sample the entire *rotation space*. With additional degrees of freedom, such as size distribution and size-dependent unusual rocking curves, one needs to have a dense and nearly-unifrom sampling in the *{size-rotation} or even in a higher-dimensional space*.

While helpful for selective suppression of coherent signatures in *partially-coherent crystallography with discrete* scattering patterns, such averaging can be detrimental to *coherent single-particle reconstructions with continuous* scattering patterns. One man's "noise" is another man's data.

## 4. Misconception of Diffraction Theory for volumetric imaging

Diffraction Theory is an *established material-independent* approximation of electromagnetic scattering for a *planar* object (aperture). It is the result of extensive theoretical and experimental studies to find the conditions for its (practical) validity.

Scattering from a 3D object can also be modeled with the same mathematical formula (applied to the 2D projection of the 3D density along the X-ray), if the Geometrical model holds. In the case of general 3D objects or even crystals, there are material-dependent and loosely-known constraints for this approximation. Referring to 3D scattering as "Diffraction Theory" in the context of pushing the boundaries of protein crystallography is not just a matter of terminology. It is an *out-of-context referral to an established approximation* and causing the misleading impression of having a theoretical foundation.

Despite *mathematical similarities* (2D Fourier transform of a projected density), scattering from a planar aperture, from a protein crystal, from a shell, and from a volume have very *different optical constraints*. Care must be taken in the use of the term Diffraction Theory and associated terms (lens-less imaging, Fresnel diffraction …) and mixing 2D (planar or morphological) problems with 3D problems. Care must also be taken in using the Geometrical model to interpret the scattering from a damaged protein crystal and to infer the dynamics of damage.

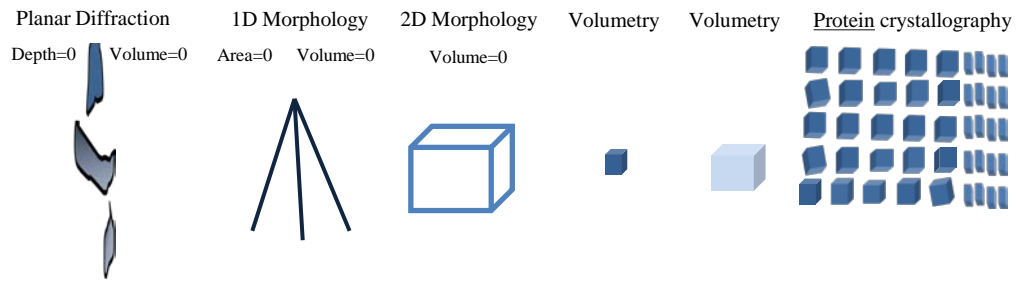

Fig. 8. Different regimes of light scattering with different constraints (on transverse size, minimum transverse feature size, longitudinal depth, index contrast …), yet all formulated with 2D Fourier transform of a projected density. *Mathematical similarity* with Fraunhofer diffraction should not overshadow *different optical constraints*. Self-consistency of the Geometrical model for a 3D object requires negligible projected density or optical path length. It can be obtained by a shell with no volume (Second and Third objects from Left in this Figure) or in a small low-density object (Second from right in this Figure) or a very small dense object (Third from right in this Figure).

The limits of the Geometrical model for volumetry are determined by the optical interaction in the sample (dependent on material properties) and not the material-independent "resolution" of detection estimated with Diffraction Theory or similar approximations. An approximate model can find the conditions for its self-consistency, but not the regime of its validity. Scattering from a carbon sphere and a gold sphere of the same size at the same wavelength look very similar, because of the pronounced signature of morphology and the symmetry of the object. In many common cases in the soft X-ray regime, however, the Geometrical model is not directly applicable for volumetric reconstruction of an object made of gold. This is a fundamental issue related to optical distortions and beyond the practical consideration of signal to noise ratio. More details about Diffraction Theory can be found in Appendix D.

## 5. Discussions and open questions

### 5.1. A closer look at nontrivial coherence in protein crystallography

The data generated in studies of imperfections in protein crystals seem to be mostly interpreted using the Geometrical model, whereas the validity of the Geometrical model itself is a function of the level of such imperfections, as shown in Fig. 1. The theory of mosaic blocks dissociates the statistical partial incoherence into deterministic fully coherent model subunits with fully-incoherent (intensity-wise) superposition. This *model* mosaic block is different from, yet related to *physical* imperfections in crystal. Both have the same optical signature, in the same way that a physical electric circuit and its reduced equivalent model have the same electrical signatures (despite magnetic, geometric … differences; and a model corresponding to many physical systems).

Addressing the exact meanings and interpretations of terms such as incoherence or mosaicity and the consistency of their use (or their concepts in the first place) is beyond the scope of this contribution. However, we compile a brief list of some reported considerations for the clarification of the context.

The incoherence associated with scattering from a protein crystal originates from both the crystal (static geometrical faults and dynamic thermal vibrations) and light (spatial and spectral spread). The term mosaicity may refer to just the geometrical component or its combinations with other factors. The mosaicity values determined from rocking-curve measurements are systematically smaller than those estimated as a fit parameter [26] by crystallography data processing programs [14,15].

In materials science, crystal imperfections in the form of *dislocations* are characterized in terms of the two basic types of *edge* and *screw* dislocation. Such patterns have also been reported for protein crystals ([27] and its citations and also [28]).

The estimated temperature factor (Debye-Waller- or B-factor) models not only the dephasing role of the temperature, but also suppresses discretization errors of Fourier analysis on a discrete grid. While efficient and elegant to utilize in data analysis [29], such additional roles of the B-factor in compensation for miscellaneous numerical errors make its physical interpretations and its use in more rigorous models [23] difficult.

### 5.2. On enhanced electron densities on surfaces

Classical formulation of scattering off symmetric homogeneous objects such as spheres shows partial light propagation in the form of *resonant* surface waves [more details in Section 2.3 of *Part 2*]. Such enhanced classical *oscillations* of the electron cloud do not change the *average position* of the electron cloud. However, with the Geometrical model, these two phenomena (enhanced illumination on the surface and enhanced electron density) are not distinguishable.

We suggest further inquiries into the observed enhanced electron density of the hydration shell surrounding biomolecules [30]. Enhanced *internal* illumination and enhanced electron density may both have potential partial contributions to this phenomenon. *Global* surface waves are avoided in protein crystallography by limiting the illumination to the middle of the *object*. Individual mosaic blocks or even unit cells, however, are likely to experience such effects *locally*. For guided waves in *2D crystals*, a uni-directional flow of the *total power* across unit cells with a single optical *mode* (despite counter-propagating electric field components) is achieved by forming power vortices [31]. Such flux vortices, if any, appear as spurious enhanced electron density with the Geometrical model.

*5.3. Protein crystallography is not a self-contained optical imaging modality*

An optical imaging modality (optical microscopy) is a self-contained measurement system; i.e., the generation and the interpretation of results are two independent tasks. Protein crystallography is *not* a self-contained *optical* imaging modality. Protein crystallography is an advanced tool of analytical chemistry to use nontrivial measured optical information along with meaningful elegant use of extensive level of *a-priori* optical information and chemical information (in real- and scattering-space) to provide additional insight into a question in chemistry.

*Volumetric* coherent diffractive imaging, as a self-contained optical imaging modality (independent of optical and chemical *a-priori* information), cannot consider protein crystallography as an example. Said differently, volumetric coherent diffractive imaging can also develop meaningful types and ways of enforcing *a-priori* information to achieve more conclusive results. It is also noted that the common phase retrieval of protein crystallography utilizes the information from two systems; i.e., a combination of phase retrieval and "interferometry" in more familiar optical terms. Even after using the highest level of common *a-priori* information (phase retrieval with Molecular Replacement), one may still need to employ unconventional *a-priori* chemical information [32] to find meaningful results [33].

An important question is the minimum "acceptable" disorder in a crystal so that it can be "ideally imperfect"; i.e., the Min value in Fig. 1. It seems that even if the disorder is below the required optical minimum $Min_{optical}$ (to make the Geometrical model applicable), the additional imposed constraints of protein crystallography data processing force the solution to the right one. How far one can go below $Min_{optical}$, depends not only on the extra information compensating for "distorted" optical data, but also on the subjective patience of an expert user to enforce such additional information. This seems to be the preferred pragmatic choice, as opposed to consideration of a more accurate scattering model (for conclusiveness, not convenience).

Model Building and Refinement fit an estimated 3D density map with known expected residues (amino acids, water, …) and impose different constraints in a systematic way, as shown in the block diagram of Figure 12-4 in [1]. A *modified* version of the block diagram in [1] with emphasized explicit role of scattering and rocking curve models seems to be close to what depicted in Fig. 9.

A crucial aspect of Model-Building/Refinement is achieving the maximum likelihood between the reconstructed diffraction intensities $\hat{I}_{(h,k,l)}$ and corresponding measured values *after integration* $I_{(h,k,l)}$. Systematic authentic deviation of $I_{(h,k,l),n}$ values of different snapshots from the average $I_{(h,k,l)}$, however, seems to have a less noticed fate. *A-priori* information and nontrivial effect of averaging on complicated rocking curves (Sec. 3.3.3) serve to suppress not only the *noise* component $I^{noise}$ of measured intensities (mechanical, optical, optoelectronic, and electronic offsets and fluctuations), but also the authentic optical information $\delta I$ corresponding to the error of the Geometrical model.

Robustness to deviation of optical scattering (from what predicted by the Geometrical model) and also high-resolution details of density map along a covalent bond can also imply that the scattering from a *slightly* damaged crystal may be tolerated in 3D reconstruction. A slightly-damaged crystal may be identified using a purely optical signature such as (modified) rocking curve, while still being suitable for the entire image reconstruction pipeline. This can mean *robustness* in the context of analytical chemistry and *error* in the context of self-contained coherent optical imaging.

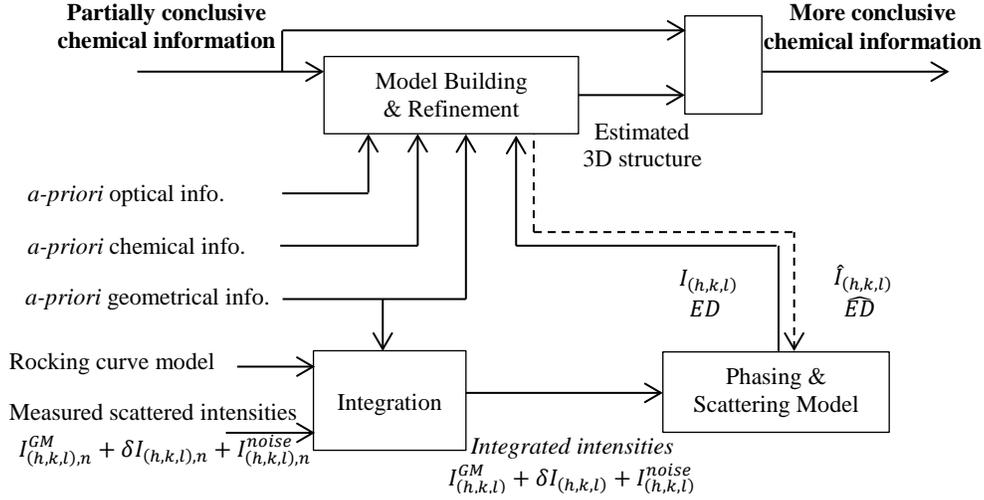

Fig. 9. Block diagram of data processing of conventional protein crystallography with emphasis on the roles of *a-priori* optical, chemical, and geometrical information, optical scattering model, and rocking curve model. Convergence of $I_{(h,k,l)}$ to $I_{(h,k,l)}^{GM}$ is facilitated when averaged $\delta I_{(h,k,l)}$ (error of the Geometrical model) has the same statistics as $I_{(h,k,l)}^{noise}$. It seems possible by averaging a lot of data and/or specific considerations for the rocking curve model and/or the scattering model (Section 3.3.3). The real-space 3D structure of electron density $ED$ and the diffraction intensities $I_{(h,k,l)}$ undergo Refinements along with Model Building. The entire data processing scheme can make 3D structure estimation possible, even with some level of error of the (Geometrical) scattering model. *A-priori* optical information includes fitting to the library of scattering patterns and form factors. *A-priori* chemical information includes general (not sample-specific) chemical information, such as molecular dynamics calculations to improve fitting. *A-priori* geometrical information includes finite and specific types of 3D lattices, the right chirality of an $\alpha$-helix, possibility and ways of handling twinning among other things.

Optical expertise can be an important asset while contributing to a crystallographic project. However, it does not enable (entitle) one to either validate or reject a crystallographic reconstruction, as one does not have the scientific *jurisdiction* in the first place (and the uniqueness of inverse X-ray scattering problem with a truncated spectrum is questionable; *Part 2*: Sec. 7.5).

*5.4. From protein crystallography to single-particle imaging*

The materials presented here in *Part 1* for clarification of optical scattering in protein crystallography and also those in *Part 2* of this contribution can be summarized, as tabulated below in Table 3.

**Table 2: The nontrivial journey from *Protein Crystallography* to *Single Particle Imaging* in a nutshell**

|  | Protein crystallography | Single particle imaging | Reference |
|---|---|---|---|
| **Scattering pattern** | Discrete | Continuous |  |
| **Sample handling (delivery)** | Single secured rotating crystal | Random orientation, size, position |  |
| **Illumination gradients** | Robust | Vulnerable | Part 1, App. B |
| **Optical *a-priori* info** | Significant | Perceived unnecessary | Part 1, Sec. 5.3 |
| **Non-optical *a-priori* info** | Significant | Perceived unnecessary | Part 1, Sec. 5.3 |
| **Self-contained imaging** | No | Perceived possible | Part 1, Sec. 5.3 |
| **Scientific jurisdiction** | Chemistry | Optics | Part 1, Sec. 5.3 |
| **Modal illumination** | Yes | No | Part 1 App. C & Part 2, Sec. 4.2 |
| **Surface modes** | Negligible | Vulnerable | Part 2, Sec. 2.3, 7.3 |
| **Non-classical scattering** | No problem | Nontrivial signatures | Part 2, Sec. 5 |
| **Quantum confinement** | Nearly irrelevant | Relevant (core-shell …) | Part 2, Sec. 5.2, 6 |
| **Meaning/Type of "density map"** | Fixed, real | Diverse, complex | Part 2, Sec. 6 |
| **Phase retrieval ($q_{Max} = \infty$)** | Highly overdetermined, 2 measurements | Overdetermined, 1 measurement | Part 2, Sec. 6.6, 7.5 |
| **Highest resolution ($q_{Max}$)** | "Absolute" (fit to residues) | Coupled to missing orders | Part 2, Sec. 7.5 |

## 6. Conclusions

The efforts seeking to push the boundaries of X-ray *protein* crystallography towards single-particle imaging have already developed considerable expertise and achievements in the experimental and statistical parts. Theoretical considerations and hence the design/optimization/interpretation of experiments, however, have been mostly focused on the topic of "damage" and on *scientific* applications. Given the diversity of samples and illumination scenarios, there is a need for more rigorous treatment of *light propagation* in less-explored regimes of X-ray optics.

Compared to light-scattering experiments at optical (VIS/NIR) wavelengths, single-particle imaging with X-rays has a more ambitious goal (volumetric imaging rather than cross section calculation), yet a more primitive model of light propagation. This inconsistency seems to have been partially capitalized on the unparalleled success of protein crystallography and a common perception of the technology behind it in "*uniform coherent* illumination of the *entire electron* density because of weak interaction with X-rays". This notion becomes more accurate and productive by enhancing its currently-blurred boundaries of validity. This fuzzy boundary seems to have been passed by some nanocrystals and single-particles already.

In conventional protein crystallography, deviations of measured scattered intensities (of individual snapshots and before integration) from the ones expected can be averaged out, and the remaining information seems sufficient for 3D reconstruction thanks to strong and meaningful imposed constraints. Despite systematic procedures for Model Building and Refinement, identification of potential spurious reconstructions may sometimes come down to the intuition of expert users [34]. These sparse regions of efficient optical models (for highest-resolution *optics-related* imaging with demanding applications) have much room for both fundamental and applied optical studies. Revisiting "failed" cases in a cause-oriented rather than result-oriented approach [35] by finding patterns such as low-mosaicity crystals [16] may be worth trying.

Such optical studies would potentially bring more objectivity and new perspectives to the evaluation of crystallographic reconstructions. However, misinterpretation of protein crystallography as a self-contained optical imaging (and an example for *ab-initio volumetric* coherent diffractive imaging) can further complicate the context of such unorthodox research directions. It might lead to the awkward situation of "scientifically acceptable, yet technologically questionable" results, with swapped perception of Science and Technology and blurred boundaries between jurisdictions of different disciplines (chemistry vs. optics) in validation of "the results".

The (over-) simplified picture of light scattering in protein crystallography manuals is strength of the technology in user-friendliness and also modular internal design (software development while keeping a friendly distance with Maxwell's equations). This picture, however, should not be misinterpreted as a foundation for advancing the technology or even a straightforward study of crystal disorders. In transition from protein crystallography to single particle volumetry, more challenges exist, which will be addressed in *Part 2*.

Pessimistic consideration of the challenges ahead of single particle imaging as inevitable show-stoppers can be as far from objectivity as an optimistic view relying on the oversimplified outlook of protein crystallography. In retrospect, Kirchhof's Diffraction Theory was even found to be inherently inconsistent. However, it soon became an established technique after conducting rigorous theoretical and experimental studies to find its scope of validity (reasonable practical accuracy). Many "fundamental physical limits" can be broken or bypassed when they are first identified and defined, with recent familiar examples being optical nanoscopy [36] and magnetic hard disk technology [37]. Such a route is also open to X-ray imaging, when optimism is completed with critical objectivity.

**Appendices**

*Appendix A: Contexts of applications of X-ray scattering models*

Some major contexts of application of X-ray scattering models are as follows:

1. The simplicity of the Geometrical model for inverse problems has had a pivotal role in what *protein crystallography* has accomplished. Significant knowledge has also been developed in the community how to use the Geometrical model more efficiently [1].

In the context of imaging applications, not only the formulations, but also the terminology and the mindset are based on the Geometrical model. The key term *reciprocal space* is used interchangeably to refer either to the scattering coordinate $\vec{q} = \vec{k}_{out} - \vec{k}_{in}$ or to the coordinate within which the Fourier transform of the scattering object is modeled $F(\vec{Q}) = \iiint f(\vec{r}) e^{-i2\pi \vec{Q}\cdot\vec{r}} dV_{\vec{r}}$. When the Geometrical model is not applicable, $I(\vec{q}) \neq |F(\vec{Q} = \vec{q})|^2$, and terms such as "reciprocal space function" or "measurement in the reciprocal space" may develop faulty references.

2. The Electromagnetic model, used in the context of *perfect* (semiconductor) crystals with a periodic index of refraction, has been formulated in different ways. The term "Dynamical Theory" is used in the X-ray community to refer to such models [10,38,39]. The main feature of light propagation in perfect crystals is the generation of two *beams* (yet many more k-vectors) inside the crystal. An intuitive *interpretation* of these rigorous results may be via the Hybrid model. Qualitatively, the two *beams* undergo ray-optical refractions and reflections at interfaces (facets of perfect crystal). This includes bending of rays along edges in a *slightly* deformed crystal. At the end, the coherent superposition of the beams gives rise to a local illumination, and a scattering from this illuminated density.

Electromagnetic model is usually not used for imaging. Mathematically, it is much more difficult to be used for inverse (and even forward) problems, compared to the common Geometrical model. Physically, the crystals requiring such a level of accuracy for modeling light propagation have very simple unit cells. The perfect crystals studied with the Electromagnetic model have lattice constants on the order of $\sim 1\text{Å}$ (covalent bonds), as opposed to protein crystals with lattice constants almost 2 orders of magnitudes larger.

Weak light-matter interaction with (hard) X rays does play an important role in effectively generating only two *beams* (in most, not all cases). However, discarding non-trivial implications of the two-*beam* model and reducing it to the Geometrical model for an arbitrary object (especially with soft X-rays) is an exaggeration of "weak light matter interaction".

The core of electromagnetic formulation of X-ray scattering in 3D crystals was developed by Ewald [39], before similar formalisms be used in solid state physics; named after Bloch; and borrowed back in optics as *optical Bloch waves*. Ewald's Habilitation thesis, however, was perceived as a beautiful mathematical construct that will never find practical applications [40]. Electromagnetic models got practical significance with the two waves of high-purity growth of semiconductor crystals and the associated market, and then the availability of high-coherence synchrotron radiation sources. Will nanocrystals or single particles initiate a third revival?

3. Electromagnetic theory is also used along with a *macroscopic* refractive index of *bulk* in the analysis and design of X-ray optical elements, in which case, it is usually reduced to Fresnel coefficients of transmission and reflection for a plane boundary.

4. Coherent scattering from individual identical objects at different orientations and the incoherent addition of these patterns is the basis of *Small-Angle X-ray Scattering* or SAXS [41]. It can be done either in liquid phase (solution scattering) or solid phase (powder pattern). Most SAXS studies also use the Geometrical model.

The prediction of the Geometrical model for the asymptotic behavior of a *classical* SAXS profile at large values of scattering vector $|\vec{q}| = |\vec{k}_{out} - \vec{k}_{in}|$ is a fall-off as $1/q^4$, known as Prod's law. Experimental observation of smaller slopes has been attributed to the known failure of the Geometrical model at grazing incidence, where significant (or even total external) reflection can happen [8,42]. In the forward problem, the Hybrid model attributes the steepness of the asymptotic curve to the surface morphology; $\sim 1/q^4$ for a volume with a smooth surface, $1/q^4$ to $1/q^3$ for a rough surface, and $1/q^3$ to $1/q^1$ for a mass fractal, as shown in Fig. 10. In other words, the slope is correlated with the fractal dimension of the surface [43]. In *Grazing incidence SAXS* studies [44], an unknown density in a thin film is probed in the reflection geometry and reconstructed with the perturbative Hybrid model.

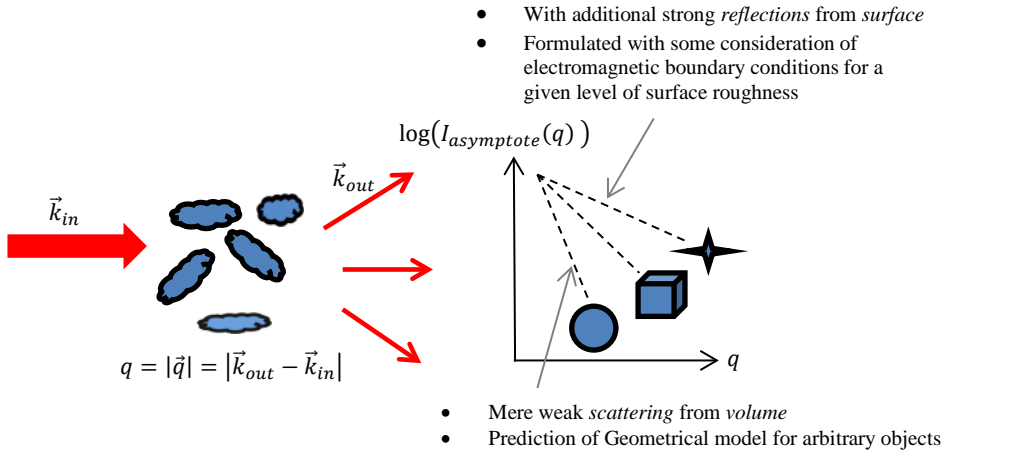

Fig. 10. (Left) Schematic of a Small-Angle X-ray Scattering (SAXS) experiment, in which coherent scattering from identical objects at different orientations are added incoherently and averaged over all orientations. (Right) Prediction of the Geometrical model for the asymptotic behavior of the SAXS intensity vs. scattering vector is only valid for homogeneous objects with smooth surfaces. Explicit

consideration of electromagnetic boundary conditions predicts smaller slopes for objects with rough surfaces.

Using the information of a SAXS measurement for reconstruction with the Geometrical model and observing considerable deviation from $1/q^4$ at large scattering angles of the same profile (in the classical regime) are in-principle inconsistent. At high non-classical resolutions, the SAXS profiles corresponding to homogeneous objects with different surface morphologies (Fig. 10) will be pushed down by the same amount (corresponding to form factor). Despite different absolute slopes (compared to the classical case), the *relative* slopes between two given morphologies remain the same as that in the classical case.

5. Volumetry of the stress pattern in an otherwise-perfect crystal is a special case, in which the geometry may be estimated, and the (stress) density is sought. Knowing the geometry can make the Hybrid model [45,46] a suitable choice for the inverse problem. An integral-form of the Electromagnetic model [47] has been used recently in the forward problem.

*Appendix B: Implications of non-uniform illumination*

In plane-wave incidence on a finite-thickness object, coherent interference of multiple reflections and transmissions from different boundaries, as shown in Fig. 11, give rise to a local electric field and an *internal* illumination pattern, which may be considerably different from the *external* field (incident plane wave).

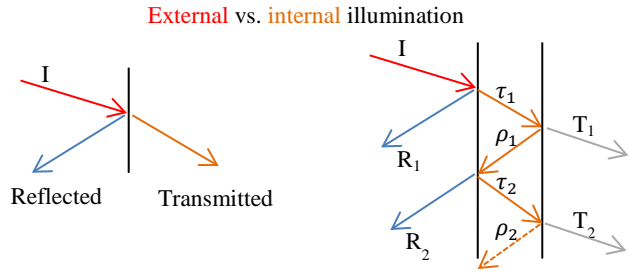

Fig. 11. (Left) An incident plane wave generates a reflected and a transmitted beam upon incidence on the flat boundary of a homogeneous semi-infinite medium. (Right) In an object with finite thickness, each plane wave component undergoes similar transmissions and reflections upon incidence on a surface. The net coherent interference of all plane wave components gives rise to an *internal* illumination pattern. Even with uniform *external* excitation (plane wave), the internal illumination can be non-uniform.

At normal incidence and with flat boundaries, the difference between external and internal illumination pattern may be discarded. However, by approaching the grazing incidence regime, reflections increase up to 100% and lead to zero transmitted beam. Furthermore, resonances inside the volume (and for closed objects also on the surface) can be formed.

An important question in conventional protein crystallography and also in experiments with nanocrystals and single particles is whether the imaged sample has been modified by radiation damage, and if so, what the mechanism is. Many such analyses use the Geometrical model, not only for crystals, but also for a damaged sample. It is insightful to note the prediction of the more accurate Electromagnetic model for the simplest illumination scenario. The counter-intuitive result of non-uniform internal illumination pattern is the beginning of (static threshold-based) damage at the back side of an object, which is embedded in a lower-index medium and illuminated with an increasing level of laser light (Fig. 12). This classic example of Fresnel reflections demonstrates a *localized* damage mechanism, which originates from constructive interference at the back boundary ($\tau_1$ and $\rho_1$ in Fig. 11) and destructive interference at the front boundary ($\rho_1$ and $\tau_2$ in Fig. 11).

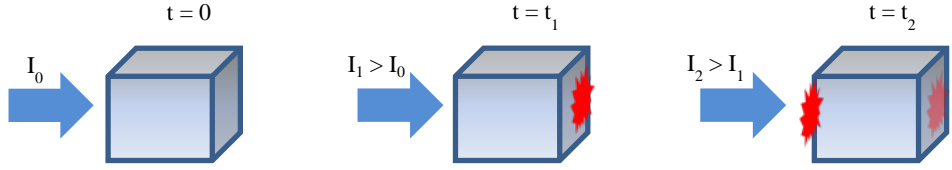

Fig. 12. Counter-intuitive evolution of static damage in a homogeneous object embedded in a lower-index medium and illuminated with a laser source with increasing level of intensity: Constructive interference at the back side results in a higher *internal* field amplitude and reaching the damage threshold earlier, compared to the front side with destructive interferences and a lower *internal* field amplitude: (Left) Initially and with an intensity below the damage threshold, a laser light simply illuminates an object, yet by creating a higher electric field at the back side, as required by boundary conditions. (Middle) Later and with an increased level of laser intensity, the internal intensity at the back side exceeds the damage threshold. With the same level of external illumination intensity, the internal intensity at the front side is still below the damage threshold. (Right) Further increase of the external illumination intensity causes the intensity at the front side to reach the damage threshold at a later stage.

Fig. 13 illustrates the robustness of protein crystallography (with incoherent superposition of patterns from different mosaic blocks) and the relative vulnerability of coherent single particle imaging to (unaccounted) gradients of the amplitude and phase of the local field. Such non-uniform internal illuminations can be formed even with uniform external illumination.

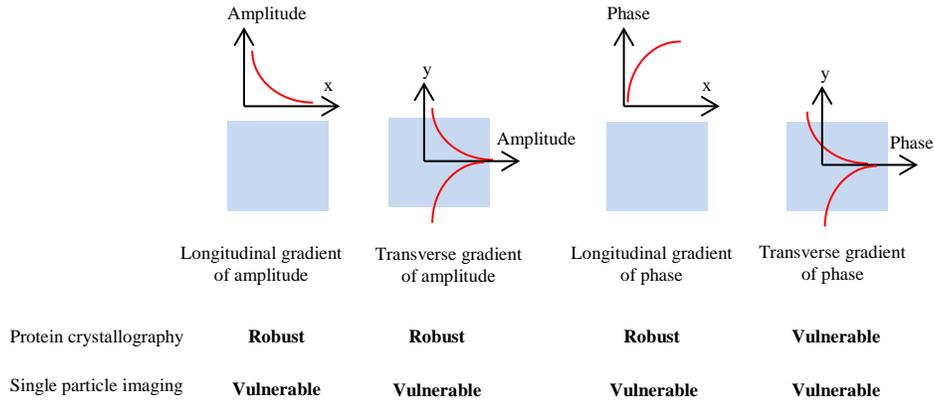

|  | Longitudinal gradient of amplitude | Transverse gradient of amplitude | Longitudinal gradient of phase | Transverse gradient of phase |
|---|---|---|---|---|
| Protein crystallography | **Robust** | **Robust** | **Robust** | **Vulnerable** |
| Single particle imaging | **Vulnerable** | **Vulnerable** | **Vulnerable** | **Vulnerable** |

Fig. 13. Robustness of crystallography (with incoherent blocks), and relative vulnerability of (coherent) single particle imaging to variations of the local field in phase and amplitude along the longitudinal and transverse directions (parallel and perpendicular to the incident X-ray).

*Appendix C: More on Electromagnetic description of X-rays in crystals*

*C.1. A closer look at perfect crystals*

The two *beams* generated in a perfect crystal represent the two dominant Floquet components of a single spatial wavepacket (Bloch mode). By changing the wavelength, the incidence angle, or the polarization, the properties of the mode change. The generation of the two beams is *not* a polarization effect (birefringence). With two polarizations, one would excite two modes and hence generate four beams (two for each polarization).

An ordinary plane wave in free space is described by its $k$ vector and a transverse electric field $E$. Mathematically, a *beam* represents a similar concept when $k$ and $E$ are replaced with $S$ (power flow or Poynting vector) and $D$ (Displacement vector). With only two Fourier components $D = D_o + D_h$ (defined in Table 2 and Fig. 14), the displacement vector is then modulated by many Fourier components of the refractive index, and causes many additional weak scattered $(k, E)$ plane waves. Table 2 compares and contrasts the numbers of Fourier

(plane-wave expansion) components of different vectors associated with an electromagnetic field in different media and with different models.

If the size of the beam incident on the crystal is large enough (to approximate a plane wave) and small enough (to resolve slightly divergent beams and to avoid illumination of edges and generation of surface effects), the beams can be separated and monitored upon exiting the crystal. This was done by Authier and gave more objective meaning to the *beams* [40].

The two-beam approximation with "weak" indices also holds at much larger index contrasts *compared to X-ray indices*; in the optical regime with indices greater than one. The transverse beam patterns can be highly structured and also changing down the path of propagation. Nevertheless (contrary to diffraction in free-space propagation), the beam width remains almost unchanged. These conclusions have been obtained with the scalar field approximation and discarding complex wavevectors. There are no (direct) restrictions on the number of Floquet components of the electric or displacement vectors (contrary to the common 2-beam or less common 3-beam and N-beam approximations of X-ray "Dynamical Theory"). The main approximation is considering a locally-quadratic dispersion surface. The results [48] have been validated with the rigorous modal theory of 3D crystals with explicit consideration of boundary conditions at infinity [49].

**Table 3: Number of components of the four electromagnetic vectors in different media with plane-wave excitation. The scattering pattern is formed with many *k*-vectors and their corresponding perpendicular electric fields *E*. Inside the crystal, the *beams* are best described by the Displacement vectors *D*, which are transverse to the power flow (Poynting) vectors *S*. The acronyms GM and EM stand for Geometrical Model and Electromagnetic Model, respectively.**

| Medium | E | D | k | S | $E = f(D)$ |
|---|---|---|---|---|---|
| Free-space | 1 | 1 | 1 | 1 | $E = D/\epsilon_0$ |
| Bulk matter | 1 | 1 | 1 | 1 | $E = D/\epsilon_0\epsilon_r$ |
| Xtl (large $\delta n$) | Many | Many | Many | Many | $E(r) = D(r)/\epsilon_0\epsilon_r(r)$ |
| Xtl ($\delta n \ll 1$), GM | Many | **1** | Many | **1** | $E(r) = D_o/\epsilon_0\epsilon_r(r)$ $\sim D_o/\epsilon_0 + (D_o/\epsilon_0)(1 - \epsilon_r(r))$ |
| Xtl ($\delta n \ll 1$), EM | Many | **2** | Many | **2** | $E(r) = (D_o + D_h)/\epsilon_0\epsilon_r(r)$ |

The reduction of the 2-*beam* model to a single-*beam* one seems to be consistent with the picture drawn by the Geometrical model. However, this single *beam* is indeed the reduced form of a *crystal mode* as internal illumination, excited by an external plane wave. The modal propagation of light in these two cases can be better understood by comparing them with the more familiar case of guided mode in an optical fiber, which is split into two branches. Plane-wave-like propagation in crystals or fibers is in principle because of the existence and the special form of the notion of *volume*-propagating mode; a luxury that does not exist in arbitrary single particle imaging. On the contrary, a resonant *surface* mode may be excited as an artifact enhancing morphological signatures rather than illuminating the volume.

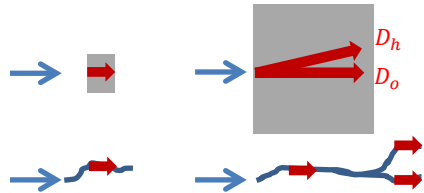

Fig. 14. Propagation of plane-wave-like *optical mode* in the volume of a crystal (Top) can be compared with the propagation of modes in the volume of (an ordinary or a bifurcated) optical fiber (Bottom). The two intense beams have been signified by their Displacement vectors $D_o$ and $D_h$, as used in Table 2. The existence and the special form of optical modes and also appropriate coupling of external illumination to such modes is the reason for a plane-wave-like internal illumination in a small crystal. In single particle imaging of small objects, such modes with uniform propagation in volume do not exist. On the contrary, resonance modes circulating on the surface can be excited, which further enhance the signatures of morphology. The fibers are depicted with twisted shape to emphasize the propagation of *guided optical mode* in the medium, rather than the incident beam simply sweeping the medium.

*C.2. Crystals with minor incoherence*

Even in the absence of geometrical imperfections, non-zero thermal energy introduces different vibrational phases across the unit cells, and the notion of "perfect crystal" can be challenged. For very small levels of incoherence in a crystal, the 2-*beam* model of scattering is still valid, with the consideration of the beams being bent by deformations. The term "Ray-tracing" (not to be mistaken with geometrical optics) is used in the context of X-ray scattering for the case of bent beams, along with the scattering they cause. Following bent rays along edges is similar to the Geometrical Theory of Diffraction in optics [50]. However, considering rays as sources of diffraction is similar to the Continuous Coupled-Mode theory in optics [51].

Ray-tracing (minor incoherence) is not applicable to mosaic blocks (nearly full incoherence). As the level of incoherence increases, the assumptions of the 2-*beam* model fail, and one needs to consider "inter-branch" scatterings [10]. In the terminology of (UV/VIS/NIR) optics, the situation is similar to what happens in an optical fiber/waveguide and tunnels multiple modes across a distorted single-mode waveguide. Exciting a single-mode optical waveguide indeed excites many higher order (discrete and even continuous) modes [52]. In a perfect waveguide, all such modes (required for satisfying boundary conditions) decay quickly and are not seen in the output. In a disordered waveguide, such modes are further coupled from one stage to the next one and may even make it all the way to the end and contribute to the output.

Ray tracing has been used for alternative interpretations [23] of scattering from a Gold crystal with finite cross sections [53]. We have not been able to verify if the briefly-reported results correspond to the Bragg regime (single reflection with finite thickness and infinite cross section) or multi-reflection scenarios (finite transverse size). In any case, this study seems to be one of few such approaches in the community focused on nanocrystals and single particles. The recommended $\sim 1\mu m$ upper-bound for thickness in the context of perfect crystals with stress [46] is similar to the $\sim 1\mu m$ limit reported for Si and Ge crystals with finite cross sections [21]. While the approach is insightful, care must be taken in using this specific upper bound on thickness for "coherent subunits" of protein crystals (with lattice constants almost 2 orders of magnitudes larger and with extra contribution to their incoherence) and also single particles (for which the bulk absorption can be significantly different from that of a crystal; Bromann's effect [10]).

*Appendix D: On "Diffraction Theory"*

*D.1. Terminology seeking reference*

A key term in the literature of X-ray optics is "diffraction". At optical (UV/VIS/NIR) wavelengths, this term is used to refer either to 1) paraxial scattering from a planar aperture, or 2) specific discrete scattering patterns of crystals. The far-field scattering in the former case is described by 2D Fourier transform of the aperture transmittance. In the latter case, the array of diffraction spots is known to correspond to the reciprocal-space lattice. However, relative intensities, optical bandgaps, and more properties are the subject of ongoing research [49,54,55]. On the other hand, it seems that in the context of imaging applications of X-rays, the concepts of light scattering and Fourier transform (Geometrical model) are equivalent notions, sometimes referred to as "Diffraction Theory".

As long as the context of application is limited (to hard X-ray scattering from ideally-imperfect protein crystals with appropriate illumination confined to the middle of the crystal), there may be no problem, and it may be only one (non-transparent) terminology conflict with

UV/VIS/NIR optics. However, using an over-simplified picture of scattering (for pushing the boundaries of protein crystallography) and referring to it with an (established, yet) irrelevant theory, can create misconceptions and not just terminology conflicts.

*D.2. What Diffraction Theory is*

Diffraction Theory is a scalar (paraxial) single-variable approximation of vectorial two-variable $(\vec{E}, \vec{H})$ equations of electromagnetics. The inherent inconsistency of Kirchhoff's formulation of Diffraction Theory was addressed in alternative formulations, and they were all shown to be reasonably accurate and equivalent for *geometrical* features ($d \gg \lambda$) and at distances not very close to the scattering *plane* [56]. In the transmission geometry, the electric field of the light transmitted through a *thin mask* with geometrical features is considered the product of the incident electric field and the complex transmittance of the mask. In a practical multi-element system, Diffraction Theory can be used successively to describe the net diffraction of the system. The list of elements includes thin masks, lenses, and bulk. Successive applications of Diffraction Theory, however, *cannot* be used to analyze the continuum of a 3D object. Even a thin homogeneous lens itself cannot be analyzed with 2D Diffraction Theory. Ray optics is used to approximate light-propagation in a thin lens [56].

The key idea behind Diffraction Theory is correlating electromagnetic variations inside a region to variations on the boundary enclosing it (via Green's theorem of vector Calculus). A rigorous formulation of scattering developed in the context of [*rigorously benchmarking the mainstream*] Diffraction Theory [57] can also be employed for a rigorous formulation of volumetric scattering of an object made of arbitrary homogeneous zones [58-60]. Care must be taken in interpreting "Diffraction Theory" and its scope of validity in old literature.

*D.3. What Diffraction Theory is not*

The reduction of 3D (Mie) scattering to the Fourier transform of a projected aperture in the case of a homogeneous sphere should be interpreted with care. Even this special case has been addressed with correction for surface waves, in the context of effective cross sections (and not *volumetry* of heterogeneous objects), and with uncertain validity: "To the extent that replacing a particle very much larger than the wavelength by an opaque planar obstacle with the same projected area is a valid approximation …" (Section 4.4.3 in [61]). Similarities of the scattering from a 3D homogeneous object under *unknown appropriate conditions* with diffraction of a corresponding 2D aperture is insightful, but is not a ground for using Diffraction Theory for *volumetry*.

A reasonable framework to *approximately* quantify the validity of the Geometrical model for volumetry, irrespective of the implications of the Diffraction Theory, is the Rayleigh-Gans approximation. It requires a small difference between a plane wave having passed through an object and one having propagated in background medium. For a homogeneous sphere with the radius $R$ and refractive index $n$, it means $4\pi R|n-1| \ll \lambda$. The combination of the 3D Rayleigh-Gans approximation (where applicable, in which case the Diffraction Theory is hardly applicable in the soft X-ray regime) and Fourier-slice theorem justifies the use of 2D Fourier transform of a projection to find the far-field scattering of a 3D object. *Mathematically*, it is similar *in form* to Fraunhofer diffraction. *Optically*, however, such projections with $4\pi R|n-1| \ll \lambda$ most likely will not meet the basic $2R \gg \lambda$ requirement of Diffraction Theory, except for extremely weak index constrasts. Even in that case, the thickness will be too large for the Diffraction Theory to hold (in its established domain of validity). In the established case of protein crystallography, one achieves a resolution distance almost equal to the wavelength ($d \sim \lambda \sim 1\text{Å}$) and very large non-paraxial scattering angles; clear violations of (the irrelevant) Diffraction Theory.